\documentclass[twocolumn]{aa} 

\usepackage{natbib}
\usepackage{amsmath}
\usepackage{graphicx}
\usepackage{txfonts}
\usepackage{textcomp}
\usepackage{booktabs}

\makeatletter
\def\@biblabel#1{}
\makeatother

\newcommand{\mjup}{M$_{\rm J}$\,}
\newcommand{\msun}{M$_\odot$\,}
\newcommand{\mstar}{M$_\star$}
\newcommand{\ms}{m\,s$^{-1}$\,}
\newcommand{\teff}{T$_{{\rm eff}}$\,}
\newcommand{\mplan}{m$_b$\,sin$i$}
\renewcommand{\cite}{\citealp}

\begin{document}

\title{{Four new planets around giant stars and the mass-metallicity correlation of planet-hosting stars.}
\thanks{Based on observations collected at La Silla - Paranal Observatory under
programs ID's 085.C-0557, 087.C.0476, 089.C-0524, 090.C-0345 and through the Chilean Telescope Time under programs
ID's CN 12A-073, CN 12B-047, CN 13A-111, CN2013B-51, CN-2014A-52, CN-15A-48 and CN-15B-25.}}

  \titlerunning{}
   \author{M. I. Jones \inst{1}
           \and J. S. Jenkins \inst{2}
           \and R. Brahm \inst{3,4}
           \and R. A. Wittenmyer\inst{5}
           \and F. Olivares E. \inst{4,6}
           \and C. H. F. Melo \inst{7}
           \and P. Rojo \inst{2}
           \and A. Jord\'an \inst{3,4}
           \and H. Drass \inst{1}
           \and R. P. Butler \inst{8}
           \and L. Wang \inst{9}}
         \institute{Center of Astro-Engineering UC, Pontificia Universidad
         Cat\'olica de Chile, Av. Vicu\~{n}a Mackenna 4860, 7820436 Macul, Santiago, Chile \\\email{mjones@aiuc.puc.cl}
         \and Departamento de Astronom\'ia, Universidad de Chile, Camino El Observatorio 1515, Las Condes, Santiago, Chile
         \and Instituto de Astrof\'isica, Facultad de F\'isica, Pontificia Universidad Cat\'olica de Chile, Av. Vicu\~{n}a Mackenna
         4860, 7820436 Macul, Santiago, Chile         
         \and Millennium Institute of Astrophysics, Santiago, Chile
         \and School of Physics and Australian Centre for Astrobiology, University of New South Wales, Sydney, NSW 2052, Australia
         \and Departamento de Ciencias Fisicas, Universidad Andres Bello, Avda. Republica 252, Santiago, Chile
         \and European Southern Observatory, Casilla 19001, Santiago, Chile
         \and Department of Terrestrial Magnetism, Carnegie Institution of Washington, 5241 Broad Branch Road, NW, Washington, DC 20015-1305, USA
         \and Key Laboratory of Optical Astronomy, National Astronomical Observatories, Chinese Academy of Sciences, A20 Datun Road, Chaoyang District, Beijing 100012, China
}

   \date{}

 
  \abstract
{Exoplanet searches have revealed interesting correlations between the stellar properties and the occurrence rate of planets. In particular, different
independent surveys have demonstrated that giant planets are preferentially found around metal-rich stars and that their fraction increases with the stellar mass.}
{During the past six years, we have conducted a radial velocity follow-up program of 166 giant stars, to detect substellar companions, 
and characterizing their orbital properties. Using this information, we aim to study the role of the stellar evolution in the orbital parameters of the 
companions, and to unveil possible correlations between the stellar properties and the occurrence rate of giant planets.  }
{We have taken multi-epoch spectra using FEROS and CHIRON for all of our targets, from which we have computed precision radial velocities and we have derived 
atmospheric and physical parameters. Additionally, velocities computed from UCLES spectra are presented here.
By studying the periodic radial velocity signals, we have detected the presence of several substellar companions. }
{We present four new planetary systems around the giant stars HIP8541, HIP74890, HIP84056 and HIP95124. 
Additionally, we study the correlation between the occurrence rate of giant planets with the stellar mass and metallicity of our targets. 
We find that giant planets are more frequent around metal-rich stars, reaching a peak in the detection of $f$ = 16.7$^{+15.5}_{-5.9}$\,\% 
around stars with [Fe/H] $\sim$ 0.35 dex. 
Similarly, we observe a positive correlation of the planet occurrence rate with the stellar mass, between \mstar $\sim$ 1.0\,-\,2.1 \msun, with a maximum of 
$f$\,=\,13.0$^{+10.1}_{-4.2}$\,\%, at \mstar = 2.1 \msun.}
{We conclude that giant planets are preferentially formed around metal-rich stars. Also, we conclude that they are more efficiently formed around more massive stars, 
in the stellar mass range of $\sim$ 1.0\,-\,2.1 \msun. These observational results confirm previous findings for solar-type and post-MS hosting stars, and 
provide further support to the core-accretion formation model.}

   \keywords{techniques: radial velocities - Planet-star interactions - (stars:) brown dwarfs}

   \maketitle
%

\section{Introduction}

Twenty years after the discovery of 51\,Peg\,b (Mayor \& Queloz \cite{MAY95}), we count more than 1600 confirmed extrasolar planets. 
In addition, there is a long list of unconfirmed systems from the {\it Kepler} mission (Borucki et al. \cite{BOR10}), adding-up more than 5000 candidate 
exoplanets\footnote{Source: http://www.exoplanets.org/} that await confirmation. 
These planetary systems have been detected around stars all across the HR diagram, in very different orbital configurations, and revealing interesting 
correlations between the stellar properties and the orbital parameters. In particular, is it now well established that there is a positive correlation between 
the stellar metallicity and the occurrence rate of giant planets (the so-called `planet-metallicity' correlation; PMC hereafter). 
The PMC has gained great acceptance in the exoplanet field, since the metal-content of the proto-planetary disk is a  key ingredient in the core-accretion model 
(Pollack et al. \cite{POL96}; Alibert et al. \cite{ALI04}; Ida \& Lin \cite{IDA04}).
This relationship was initially proposed 
by Gonzalez et al. (\cite{GON97}), and has been confirmed by subsequent studies (Santos et al. \cite{SAN01}; Fischer \& Valenti \cite{FIS05}; hereafter FIS05).
Moreover, by comparing the host-star metallicity of 20 sub-stellar companions from the literature with the metallicity distribution of the Lick sample,  
Hekker et al. (\cite{HEK07}) showed that planet-hosting stars are on average more metal rich by 0.13 $\pm$ 0.03 dex, suggesting that the PMC might also
be valid for giant stars.
However, recent works have obtained conflicting results, particularly from planet search programs focusing on post-main-sequence (MS) stars.
For instance, Pasquini et al. (\cite{PAS07}), based on a sample of 10 planet-hosting giant stars\footnote{One of those planets (HD\,122430\,b) was shown not 
be a real planet (Soto et al. \cite{SOT15}).}, showed that exoplanets around evolved stars 
are not found preferentially in metal-rich systems, arguing that the planet-metallicity correlation might be explained by an atmospheric pollution effect, due
to the ingestion of iron-rich material or metal-rich giant planets (Murray \& Chaboyer \cite{MUR02}). 
Similarly, Hekker et al. (\cite{HEK08}) showed a lack of correlation between the planet occurrence rate and the stellar metallicity, although they included 
in the analysis all of the giant stars with observable periodic radial velocity (RV) variations, instead of only including those stars with secure planets. 
Thus, it might be expected that the Hekker et al. sample 
is contaminated with non-planet-hosting variable stars. D\"ollinger et al. (\cite{DOL09}) showed that planet-hosting giant stars 
from the Tautenburg survey tend to be metal-poor. 
In contrast, based on a small sample of subgiant stars with \mstar $>$ 1.4 \msun, Johnson et al. (\cite{JOHN10}; hereafter JOHN10) found that their data are 
consistent with the 
planet-metallicity correlation observed among dwarf stars. Also, Maldonado et al. (\cite{MAL13}) showed that the planet-metallicity correlation is observed
in evolved stars with \mstar $>$ 1.5\msun, while for the lower mass stars this trend is absent.
Finally, based on a much larger sample analyzed in a homogeneous way, Reffert et al. 
(\cite{REF15}; hereafter REF15) showed that giant planets around giant stars are preferentially formed around metal-rich stars. \newline \indent
On the other hand, different RV surveys have also shown a direct correlation between the occurrence rate of giant planets and the stellar mass.
Johnson et al. (\cite{JOHN07}), claimed that there is a positive correlation between the fraction of planets and stellar mass. They showed that 
the fraction increases from $f$ = 1.8 $\pm$ 1.0 \%, for stars with \mstar $\sim$ 0.4 \msun\, to a significantly higher value of $f$ = 8.9 $\pm$ 2.9 \%, 
for stars with \mstar $\sim$ 1.6 \msun.
These results were confirmed by JOHN10, who showed that there is a linear increase in the fraction of giant planets with 
the stellar mass, characterized by $f$ = 2.5 $\pm$ 0.9 \%, for \mstar $\sim$ 0.4 \msun\, and  $f$ = 11.0 $\pm$ 2.0 \%, for \mstar $\sim$ 1.6 \msun.
In a similar study, Bowler et al. (\cite{BOW10}) showed that the fraction of giant planets hosted by stars with mass between 1.5\,-\,1.9 \msun\, 
is $f$ = 26$^{+9}_{-8}$ \%, significantly higher than the value obtained by JOHN10.
\newline \indent
In this paper we present the discovery of four giant planets around giant stars that are part of the EXPRESS ({\bf EX}o{\bf P}lanets a{\bf R}ound 
{\bf E}volved {\bf S}tar{\bf S}) radial velocity program (Jones et al. \cite{JON11}; hereafter JON11).
The minimum masses of the substellar companions range between 2.4 and 5.5 \mjup, and have orbital periods in the range 562-1560 days.
All of them have low eccentricity values $e$ $<$ 0.16. 
In addition to these planet discoveries, we present a detailed analysis of the mass-metallicity correlations of the planet-hosting and non-planet-hosting 
stars in our sample, along with studying the fraction of multiple-planet systems observed in giant stars. \newline \indent
This paper is organized as follows. Section 2 briefly describes the observations and radial velocity computation techniques. In Section 3 we summarize
the main properties of the host stars. In Section 4, we present a detailed analysis of the orbital fits and stellar activity analysis. 
In Section 5 we present a statistical analysis of the mass-metallicity correlation of our host stars. Also, we discuss about the
occurrence rate of multiple-planet systems. The summary and discussion are presented in Section 6. 

\section{Observations and RV calculation}

Since 2009 we have been monitoring a sample of 166 bright giant stars that are observable from the southern hemisphere. The selection criteria of the sample
are presented in JON11. We have been using two telescopes 
located in the Atacama desert in Chile, namely the 1.5\,m telescope at the Cerro Tololo Inter-American Observatory
and the 2.2\,m telescope at La Silla observatory. The former was initially equipped with the fiber-fed 
echelle spectrograph (FECH), which was replaced in 2011 by CHIRON (Tokovinin et al. \cite{TOK13}), a 
much higher resolution\footnote{CHIRON delivers a maximum resolution of $\sim$ 130,000 using the narrow slit 
mode.} and more stable spectrograph. These two spectrographs are equipped with an I$_2$ cell, which is used as a
precision wavelength reference. \newline \indent The 2.2\,m telescope is connected to the Fiber-fed Extended 
Range Optical Spectrograph (FEROS; Kaufer et al.  \cite{KAU99}) via optical fibres to stabilise the pupil entering the 
spectrograph. FEROS offers a unique observing mode, delivering a spectral resolution of $\sim$ 48,000, and does 
not require the beam be passed through an I$_2$ cell for precise wavelength calibration, using a Thorium-Argon 
gas lamp instead. \newline \indent 
We have taken several spectra for each of the stars in our sample using these 
instruments. In the case of FECH and CHIRON, we have computed precision radial velocities using the iodine 
cell method (Butler et al. \cite{BUT96}). We achieve typically a precision of $\sim$ 10-15 \ms from FECH data, and $\sim$ 5 \ms
for CHIRON. On the other hand, for FEROS spectra, we used the simultaneous calibration method (Baranne et al. \cite{BAR96}) to extract
the stellar radial velocities, reaching a typical precision of $\sim$ 5 \ms.
Details on the data reduction and RV calculations have been given in several papers
(e.g. Jones et al. \cite{JON13,JON14,JON15a,JON15b}).
In addition, we present complementary observations from the Pan-Pacific Planet Search (PPPS; Wittenmyer et al. \cite{WIT11}). 
These spectroscopic data have been taken with the UCLES spectrograph (Diego et al. \cite{DIE90}), which delivers a resolution of R $\sim$ 45,000, 
using a 1 arcsec slit.
The instrument is mounted on the 3.9\,m Anglo-Australian telescope and is also equipped with an I$_2$ cell for wavelength calibration. 
Details on the reduction procedure and RV calculations can be found in Tinney et al. (\cite{TIN01}) and Wittenmyer et al. (\cite{WIT12}).

\section{Host stars properties}

Table \ref{stellar_par} lists the stellar properties of HIP\,8541 (=\,HD\,11343 ), HIP\,74890 (=\,HD\,135760), HIP\,84056 (=\,HD\,155233) and HIP\,95124 
(=\,HD\,181342). 
The spectral type, $B-V$ color, visual magnitude, and parallax of 
these stars were taken from the $Hipparcos$ catalog (Van Leeuwen \cite{VAN07}). The atmospheric parameters (T$_{eff}$, log$g$, 
[Fe/H], vsin$i$) were computed using the MOOG code\footnote{\url{http://www.as.utexas.edu/~chris/moog.html}} (Sneden \cite{SNE73}), following the 
methodology described in JON11. The stellar mass and radius was derived by comparing the position of the star in the HR diagram 
with the evolutionary tracks from Salasnich et al. (\cite{SAL00}). A detailed description of the method is presented in Jones et al. 
(\cite{JON11,JON15b})  

\begin{table*}
\centering
\caption{Atmosperic parameters and physical properties of the host stars.\label{stellar_par}}
\begin{tabular}{lrrrr}
\hline\hline
\vspace{-0.3cm} \\
                      & HIP\,8541          & HIP\,74890       & HIP\,84056        &  HIP\,95124      \\
\hline \vspace{-0.3cm} \\
Spectral Type         &   K2III/IV         & K1III            & K1III             & K0III            \\
$B-V$ (mag)             &   1.08             & 1.05             & 1.03              & 1.02             \\
$V$ (mag)               &   7.88             & 7.05             & 6.81              & 7.55             \\
Parallax (mas)        &   5.93 $\pm$ 0.61  & 10.93 $\pm$ 0.63 & 13.31 $\pm$ 0.59  & 9.04  $\pm$ 0.61 \\
\teff (K)             &   4670 $\pm$ 100   & 4850  $\pm$ 100  & 4960  $\pm$ 100   & 5040  $\pm$ 100  \\
L (L$_\odot$)         &   25.4 $\pm$ 5.8   & 16.4  $\pm$ 2.4  & 13.45 $\pm$ 1.73  & 14.99 $\pm$ 2.46 \\
log\,g (cm\,s$^{-2}$) &   2.7  $\pm$ 0.2   & 3.1   $\pm$ 0.2  & 3.17  $\pm$ 0.2   & 3.3   $\pm$ 0.2  \\
{\rm [Fe/H]} (dex)    &  -0.15 $\pm$ 0.08  & 0.20  $\pm$ 0.13 & 0.08  $\pm$ 0.07  & 0.20  $\pm$ 0.08 \\
$v$\,sin$i$ (k\ms)    &   1.3  $\pm$ 0.9   & 2.2   $\pm$ 0.9  & 1.67  $\pm$ 0.9   & 1.90  $\pm$ 0.9  \\
M$_\star$ (\msun)     &   1.17 $\pm$ 0.28  & 1.74  $\pm$ 0.21 & 1.69  $\pm$ 0.14  & 1.89  $\pm$ 0.11 \\
R$_\star$ (R$_\odot$) &   7.83 $\pm$ 1.02  & 5.77  $\pm$ 0.53 & 5.03  $\pm$ 0.39  & 5.12  $\pm$ 0.49 \\
\vspace{-0.3cm} \\\hline\hline
\end{tabular}
\end{table*}
 
\section {Orbital parameters and activity analysis}

\subsection{HIP8541\,b \label{sec_HIP8541}}

\begin{figure}[t!]
\centering
\includegraphics[width=8cm,height=10cm,angle=270]{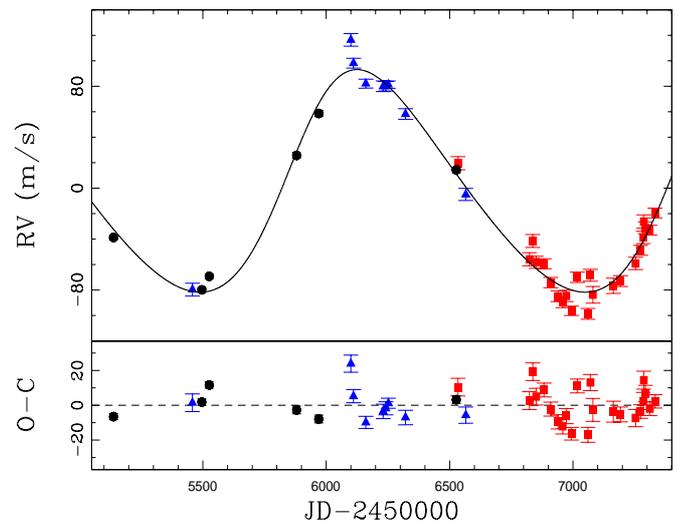}
\caption{Radial velocity measurements of HIP\,8541. The black circles, blue triangles and red squares represent 
the UCLES, FEROS and CHIRON velocities, respectively. 
The best Keplerian solution is overplotted (black solid line). The post-fit residuals are shown in the lower panel. 
\label{fig_HIP8541_vels}}
\end{figure}

We have computed a total of 36 precision RVs of HIP8541, from FEROS, CHIRON, and UCLES spectra taken between 2009 and 2015.
These velocities are listed in Table \ref{HIP8541_vels}, and are shown in Figure \ref{fig_HIP8541_vels}. 
As can be seen, there is a large RV signal with an amplitude that exceeds the instrumental uncertainties, and the RV jitter 
expected for the spectral type of this star (e.g. Sato et al. \cite{SAT05}), by an order of magnitude. 
A Lomb-Scargle (LS) periodogram (Scargle \cite{SCA82}) revealed a strong peak around $\sim$ 1600 days. Starting from this orbital period, we computed
the Keplerian solution using the Systemic Console version 2.17 (Meschiari et al. \cite{MES09}). To do this, we added a 5 \ms
error in quadrature to the internal instrumental uncertainties. This value is the typical level of RV noise induced by stellar pulsations 
in these type of giant stars (Kjeldsen \& Bedding \cite{KJE95}). 
We obtained a single-planet solution with the following parameters: P = 1560.2 $\pm$ 53.9\,$d$, \mplan = 5.5 $\pm$ 1.0 \mjup
and $e$ = 0.16 $\pm$ 0.06. The post-fit RMS is 9.1 \ms, and no significant periodicity or linear trend is observed in the RV residuals.
The RV uncertainties were computed using the Systemic bootstrap tool.
In the case of the planet mass and semi-major axis, the uncertainty was computed by error propagation, including both the 
uncertainties in the fit (from the bootstrap tool) and also the uncertainty in the stellar mass. 
The full set of orbital parameters and their corresponding uncertainties are listed in Table \ref{orb_par}. \newline   \indent
Since stellar intrinsic phenomena can mimic the presence of a sub-stellar companion (e.g. Huelamo et al. \cite{HUE08}; Figueira et al. \cite{FIG10}; 
Boisse et al. \cite{BOI11}), we examined the All Sky Automatic Survey (ASAS; Pojmanski \cite{POJ97}) $V$-band photometry and the Hipparcos photometric data of 
HIP\,8541. For both datasets, we included only the best quality data (grade A and quality flag equal to 0 and 1, respectively). 
We also filtered the ASAS data using a 3-$\sigma$ rejection method to remove outliers, which are typically due to CCD saturation. The photometric
stability of the ASAS and Hipparcos data are 0.013 and 0.012 mag, respectively. Moreover, a periodogram analysis of these two datasets show
no significant peak around the period obtained from the RV time-series. Similarly, we computed the bisector velocity span (BVS; Toner \& Gray \cite{TON88}; 
Queloz et al. \cite{QUE01}) and the full width at half maximum (FWHM) variations of the cross-correlation function (CCF), from FEROS spectra. 
None of these activity indicators show any
significant correlation with the observed RVs. Finally, we computed the S-index variations from the reversal core emission of the Ca\,{\sc ii} H and K lines, 
according to the method presented in Jenkins et al. (\cite{JEN08,JEN11}), revealing no significant correlation with the measured velocities.  

\subsection{HIP\,74890\,b}\label{sec_HIP74890}

\begin{figure}[]
\centering
\includegraphics[width=8cm,height=10cm,angle=270]{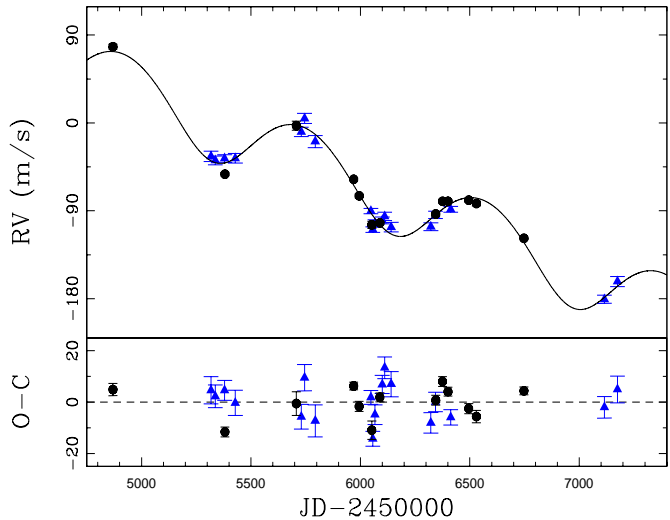}
\caption{ Radial velocity measurements of HIP\,74890. The black filled circles and blue triangles correspond to UCLES and FEROS
measurements, respectively. The solid line is the best Keplerian solution. The residuals around the fit are shown in the lower panel.
\label{fig_HIP74890_vels}}
\end{figure}

The velocity variations of HIP\,74890 are listed in Table \ref{HIP74890_vels}. The RVs were computed from FEROS and UCLES
spectra, taken between the beginning of 2009 and mid 2015. 
A detailed analysis of the RV data revealed a periodic signal, which is superimposed onto a linear trend. 
The best Keplerian fit is best explained by a giant planet with a projected mass of 2.4 $\pm$ 0.3 \mjup, in a 822-day orbit with a low
eccentricity of $e$ = 0.07 $\pm$ 0.07. A third object in the system induces a linear acceleration of -33.23 $\pm$ 1.46 m\,${\rm s^{-1 }yr}^{-1}$. 
The full set of parameters with their uncertainties are listed in Table \ref{orb_par}.
Using the Winn et al. (\cite{WINN09}) relationship, we obtained a mass and orbital distance of the outer object of 
m$_c sini > $ 7.9 \mjup and $a_c >$ 6.5 AU, respectively.  Figure \ref{fig_HIP74890_vels} shows the HIP\,74890 radial velocities. \newline \indent
The Hipparcos and ASAS photometric datasets of this stars present a stability of 0.008 mag and 0.013 mag, respectively. No significant peak is observed
in the LS periodogram of these two datasets. Similarly, the BVS analysis, CCF variations, and chromospheric activity analysis show neither 
an indication of periodic variability, nor any correlation with the radial velocities.

\begin{table*}
\centering
\caption{Orbital parameters \label{orb_par}}
\begin{tabular}{lllll}
\hline\hline 
\vspace{-0.3cm} \\
                             &   HIP8541\,{\it b}   &  HIP74890\,{\it b}   &  HIP84056\,{\it b}  &  HIP95124\,{\it b}  \\
\hline \vspace{-0.3cm} \\
P (days)                     &  1560.2 $\pm$ 53.9   &   822.3 $\pm$ 16.8   &  818.8 $\pm$ 12.1  &  562.1  $\pm$ 6.0   \\   
K (\ms)                      &  87.4 $\pm$ 6.4      &   36.5  $\pm$ 2.7    &  40.5  $\pm$ 3.1   &  46.5   $\pm$ 1.8   \\ 
$a$ (AU)                     &  2.8  $\pm$ 0.25     &   2.1   $\pm$ 0.09   &  2.0   $\pm$ 0.06   &  1.65   $\pm$ 0.04  \\
$e$                          &  0.16 $\pm$ 0.06     &   0.07  $\pm$ 0.07   &  0.04  $\pm$ 0.04   &  0.10   $\pm$ 0.07  \\
M$_{\rm P}$\,sin$i$ (\mjup)  &  5.5  $\pm$ 1.0      &   2.4   $\pm$ 0.3    &  2.6   $\pm$ 0.3    &  2.9    $\pm$ 0.2   \\
$\omega$ (deg)               &  293.9 $\pm$ 15.2    &   181.9 $\pm$ 93.9   &  120.0 $\pm$ 71.9   &  311.8  $\pm$ 35.8  \\
T$_{\rm P}$-2455000          &  4346.9 $\pm$ 93.4   &   4820.4 $\pm$ 379.8 &  5282.0 $\pm$ 192.1 &  4915.5 $\pm$ 54.3  \\
$\dot{\gamma}$ (m\,s$^{-1}$yr$^{-1}$) & -\,-\,-      &   -33.23 $\pm$ 1.46  &  -\,-\,-            &  -\,-\,-            \\
$\gamma_1$ (\ms) (CHIRON)    &  58.8  $\pm$ 4.1     &    -\,-\,-           &  6.7   $\pm$ 2.3    &  24.6   $\pm$ 3.0   \\
$\gamma_2$ (\ms) (FEROS)     &  -56.8 $\pm$ 5.6     &    78.1 $\pm$ 3.6    &  3.6   $\pm$ 2.7    &  -1.4   $\pm$ 2.6   \\  
$\gamma_3$ (\ms) (UCLES)     &  -14.3 $\pm$ 5.0     &    80.3 $\pm$ 4.3    &  -\,-\,-            &  4.8    $\pm$ 5.0   \\ 
RMS (\ms)                    &  9.1                 &    6.5               &  9.9                &  7.2                \\
$\chi$$^2_{\rm red}$         &  2.4                 &    1.5               &  2.7                &  1.7                \\
\vspace{-0.3cm} \\\hline\hline
\end{tabular}
\end{table*}

\subsection{HIP84056\,b}

\begin{figure}[]
\centering
\includegraphics[width=8cm,height=10cm,angle=270]{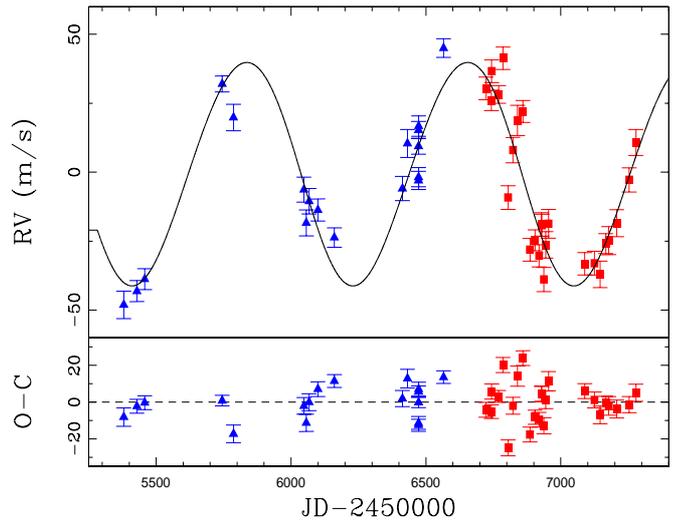}
\caption{Upper panel: Radial velocity measurements of HIP\,84056. The blue triangles and red squares correspond to FEROS and CHIRON data,
respectively.  The best Keplerian solution is overplotted (black solid line).
Lower panel: Residuals from the Keplerian fit. 
\label{fig_HIP84056_vels}}
\end{figure}

The velocity variations of HIP\,84056 are listed in Table \ref{HIP84056_vels} and Figure \ref{fig_HIP84056_vels} shows its RV curve. 
The best orbital solution leads to: P = 818.8 $\pm$ 12.1\,${\rm d}$, 
\mplan \,= 2.6 $\pm$ 0.3 and $e$ = 0.04 $\pm$ 0.04. The full orbital elements solution are listed in Table \ref{orb_par}. 
This planet was independently detected by the PPPS 
(Wittenmyer et al. \cite{WIT16}). Based on 21 RV epochs, they obtained an orbital period of 885 $\pm$ 63 days, minimum 
mass of 2.0 $\pm$ 0.5 \mjup, and eccentricity of 0.03 $\pm$ 0.2, in good agreement with our results. \newline \indent
To determine the nature of the periodic RV signal observed in HIP84056, we performed an activity analysis, as described in section 
\ref{sec_HIP8541}. We found no significant periodicity or variability of the activity indicators with the observed RVs.
Moreover, the photometric analysis of the Hipparcos data reveals a stability of 0.009 mag. Similarly, the RMS of the 
ASAS data is 0.012 mag. These results support the planet hypothesis of the periodic signal detected in the RVs.

\subsection{HIP95124\,b}

\begin{figure}[]
\centering
\includegraphics[width=7cm,height=9cm,angle=270]{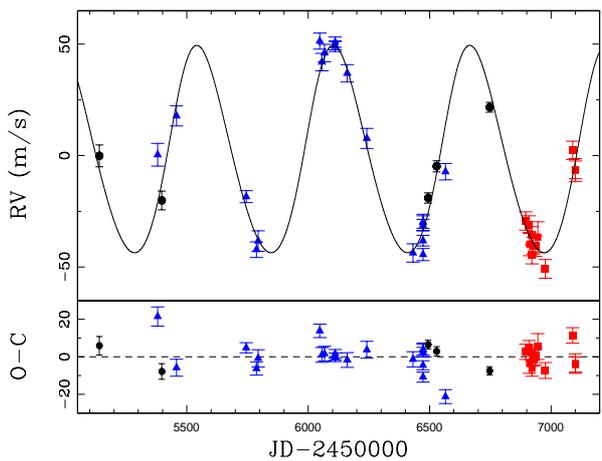}
\caption{Radial velocities of HIP\,95124. The black filled circles, blue triangles and red open circles correspond to UCLES, FEROS
and CHIRON data, respectively. 
The solid line is the best Keplerian solution. The residuals around the fit are shown in the lower panel.
\label{fig_HIP95124_vels}}
\end{figure}

Figure \ref{fig_HIP95124_vels} shows the RV variations of HIP95124.
The orbital parameters are listed in Table \ref{orb_par}.
The RV variations of HIP\,8541 are best explained by the presence of a 2.9 $\pm$ 0.2 \mjup planet, with orbital period of 
P = 562.1 $\pm$ 6.0 ${\rm d}$ and eccentricity $e$ = 0.1 $\pm$ 0.07. 
The radial velocities are also listed in Table \ref{HIP95124_vels}. As for the other stars described here, we scrutinized the Hipparcos and ASAS 
photometry to search for any signal with a period similar to that observed in the RV timeseries, finding a null result. 
Moreover, the Hipparcos and ASAS RMS is 0.007 mag and 0.013 mag. According to Hatzes (\cite{HAT02}), this photometric variability 
is well below the level to mimic the RV amplitude observed in this star. Additionally, the BVS, CCF variations, and S-index 
variations show no significant correlation with the observed radial velocities. 

\section{Preliminary statistical results of the EXPRESS project}

\begin{figure}[]
\centering
\includegraphics[width=7cm,height=9cm,angle=270]{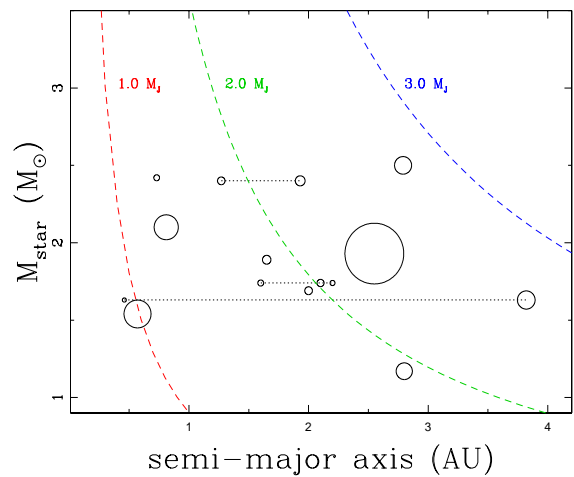}
\caption{Stellar mass versus semi-major axis of the 12 planetary systems in our sample. The size of the circles is proportional to the 
planet mass. Multi-planet systems are connected by the dotted lines. The red, green and blue dashed lines correspond to $K$ = 30 \ms, for 1, 2 and 3 \mjup planets, 
respectively.\label{fig_semiaxis_star}}
\end{figure}

After 6 years of continuous monitoring of a sample comprised by 166 giant stars, 
we have published a total of 11 substellar companions (including this work), orbiting 10 different stars.
In addition to this, using combined data of the EXPRESS and PPPS surveys, we have detected a two-planet system in a 3:5 mean-motion resonance 
(Wittenmyer et al. \cite{WIT15}) around the giant star HIP\,24275. 
Moreover, Trifonov et al. (\cite{TRI14}), recently announced the discovery of a two-planet systems around 
HIP\,5364, as part of the Lick Survey (Frink et al. \cite{FRI02}).
Since this star is part of our RV program, we have also taken several FECH and CHIRON spectra. The resulting velocities will be presented in a forthcoming 
paper (Jones et al., in preparation). \newline  \indent
In summary, a total of 15 substellar companions to 12 different stars in our sample have been confirmed, plus a number
of candidate systems that are currently being followed-up (Jones et al. in preparation). These objects have projected masses in the range 1.4\,-\,20.0 \mjup,  
and orbital periods between 89 $d$ (0.46 AU) and 2132 $d$ (3.82 AU). \newline \indent
Figure \ref{fig_semiaxis_star} shows the orbital distance versus stellar mass for these 12 systems. The red and blue dashed lines represent radial velocity 
amplitudes of $K$ = 30 \ms (assuming circular orbits), which correspond to $\sim$ 3-$\sigma$ detection limits\footnote{For FEROS and CHIRON data, the RV 
noise is dominated by stellar pulsations, that induce velocity variability of $\sim$ 5-10 \ms level in our targets. In fact, according to 
Kjeldsen \& Bedding (\cite{KJE95}), only 4 of our targets are expected to present velocity variations larger than 10 \ms. In the case of FECH data, the 
instrumental uncertainty is comparable to the stellar pulsations noise.}. It can be seen that  
we can detect planets with M$_P$ $\gtrsim$ 3.0 \mjup up to $a$ $\sim$ 3 AU (or M$_P$ $\gtrsim$ 2.5 \mjup at $a$ $\sim$ 2.5 AU) around 
stars with \mstar $\lesssim$ 2.5 \msun. For more massive stars, we can only detect such planets but at closer orbital distance ($a$ $\lesssim$ 2.5 AU for 
\mstar = 3.0 \msun).
We note that we have collected at least 15 RV epochs for each of our targets, with a typical timespan of $\sim$ 2-3 years, which allow us to efficiently detect 
periodic RV signals with $K \gtrsim$ 30 \ms and $e \lesssim$ 0.6 via periodogram analysis and visual 
inspection. Moreover, we have obtained additional data for our targets showing RV variability $\gtrsim$ 20 \ms, including those presenting linear trends. 
In fact, some of these linear trend systems are brown-dwarf candidates, with orbital periods exceeding the total
observational timespan of our survey (P $\gtrsim$ 2200 d; see Bluhm et al., submitted).
We also note that in the case of HIP\,67851\,c (Jones et al. \cite{JON15b}), we used ESO archive data to fully cover its orbital period (P = 2132 d; $a$ = 3.82 AU).

\subsection{Stellar mass and metallicity}

\begin{figure}[]
\centering
\includegraphics[width=7cm,height=9cm,angle=270]{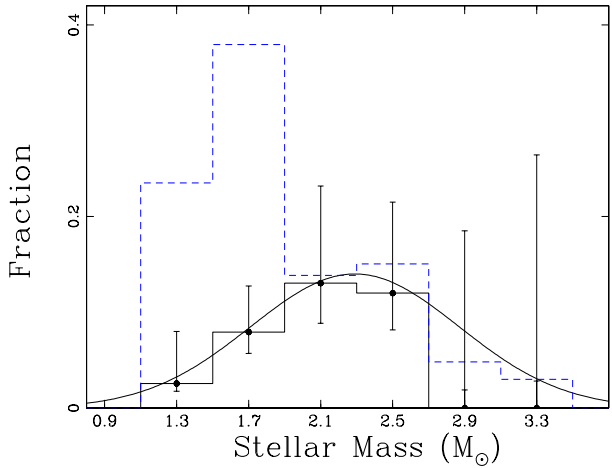}
\caption{Normalized occurrence rate versus stellar mass for EXPRESS targets with published planets. The dashed blue line corresponds to the parent sample 
distribution. The solid curve corresponds to the Gaussian fit (equation \ref{eq1}).
\label{occ_rate}}
\end{figure}

\begin{figure}[]
\centering
\includegraphics[width=7cm,height=9.5cm,angle=270]{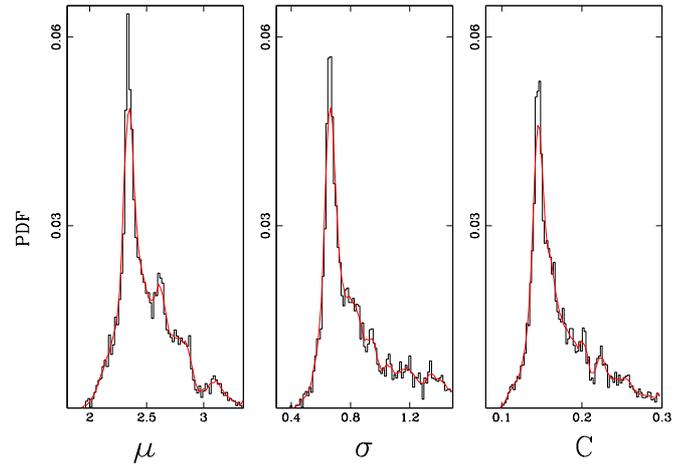}
\caption{Probability density functions for $\mu$, $\sigma$, and $C$,  obtained from a total of 10000 synthetic datasets. The red lines correspond to the smoothed
distributions.  
\label{C1_C2}}
\end{figure}

\begin{figure}[]
\centering
\includegraphics[width=7cm,height=9cm,angle=270]{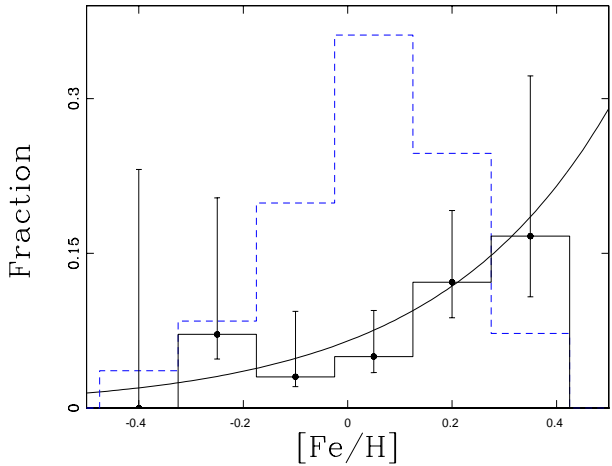}
\caption{Normalized occurrence rate versus stellar metallicity for EXPRESS targets with published planets. The dashed blue line corresponds to the parent sample 
distribution. The solid curve is our best fit to equation \ref{eq2}.
\label{occ_rate_metallicity}}
\end{figure}

\begin{figure}[]
\centering
\includegraphics[width=7cm,height=9cm,angle=270]{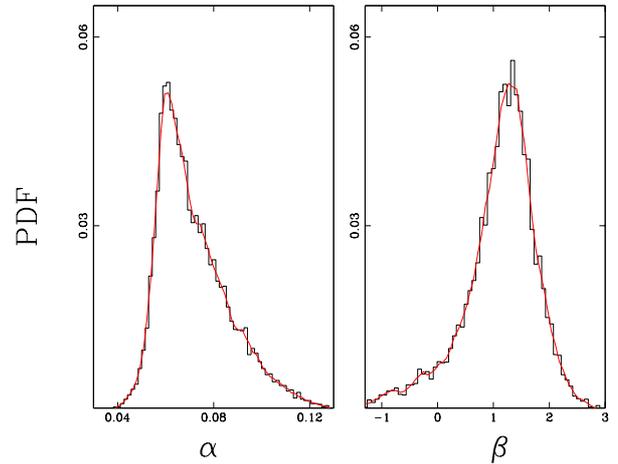}
\caption{Probability density functions for $\alpha$ and $\beta$ obtained from a total of 10000 synthetic datasets. The red lines correspond to the smoothed 
distributions.
\label{alpha_beta}}
\end{figure}

\begin{figure}[]
\centering
\includegraphics[width=8cm,height=10cm,angle=270]{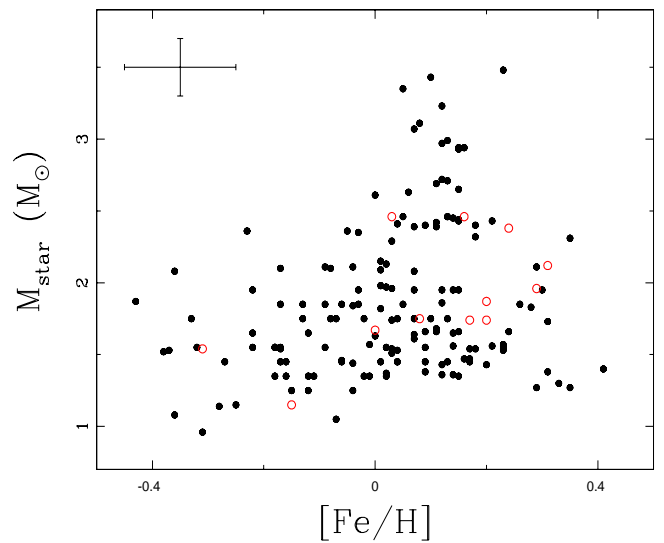}
\caption{Mass versus metallicity of the 166 giant stars in our survey. The red open circles correspond to the planet-hosting stars.
\label{mass_fe_H}}
\end{figure}

Figure \ref{occ_rate} shows a histogram of the planetary occurrence rate as a function of the stellar mass in our sample.
The bin width is 0.4 \msun and the stellar masses range from 0.9 \msun to 3.5 \msun. The uncertainties were computed according to Cameron (\cite{CAM11}), 
and correspond to 68.3\% equal-tailed confidence limits.
As can be seen, there is an increase in the detection fraction with the stellar mass, between $\sim$ 1.0\,-2.1 \msun, reaching a peak in the occurrence rate of 
$f$\,=\,13.0$^{+10.1}_{-4.2}$\,\%, at \mstar = 2.1 \msun. 
In addition, there is a sharp drop in the occurrence rate at stellar masses $\gtrsim$ 2.5 \msun. 
In fact, there are 17 stars in our survey in this mass regime, but none of them host a planet. We note that, although the observed lack of planets around these 
stars might be in part explained by the reduced RV sensitivity (see Figure \ref{fig_semiaxis_star}), all of our targets more massive than 2.5 \msun present
RV variability $\lesssim$ 15 \ms. This means that we can also discard the presence of planets with M$_P$ $\gtrsim$ 3.0 \mjup interior to $a$ $\sim$ 3 AU, otherwise we 
would expect to observe doppler-induced variability at the $\gtrsim$ 20 \ms level.\footnote{The RMS of a cosine function (circular orbit) is $\sim$ 0.71\,$K$, 
where $K$ is the semi-amplitude of the signal. Thus, for a $K$\,=\,30 \ms semi-amplitude RV signal, we expect to observe a variability (RMS) of $\sim$ 21 \ms, 
while for $K$\,=\,21 \ms  is $\sim$ 15 \ms.}
Following the REF15 results, we fitted a Gaussian function to the data, of the form:
\begin{equation}
     f(M_\star) = C\,{\rm exp}\left(\frac{-(M_\star - \mu)^2}{2\,\sigma^2}\right). 
     \label{eq1}
\end{equation}
To obtain the values of $C$, $\mu$ and $\sigma$, we generated 10000 synthetic datasets, computing the confidence limits for each realization following the 
Cameron (\cite{CAM11}) prescription. 
After fitting $C$, $\mu$ and $\sigma$ for each synthetic dataset, we end-up with a probability density distribution for each of these three parameters.
We note that we first computed $C$, and then we fixed it to compute $\mu$ and $\sigma$, restricting these two parameters to: $\mu$ $\in$ [1.5,3.0] and $\sigma$ 
$\in$ [0.0,1.5].                   
Figure \ref{C1_C2} shows our results for the three parameters. 
The red lines correspond to the smoothed distributions. 
We obtained the following values: $C$ = 0.14 $^{+0.08}_{-0.01}$, $\mu$ = 2.29 $^{+0.44}_{-0.06}$ \msun, and $\sigma$ = 0.64 $^{+0.44}_{-0.03}$ \msun. 
The parameters were derived from the maximum value and equal-tailed confidence limits of each smoothed distribution, respectively.  \newline \indent 
Despite the fact that we are dealing with low number statistics, particularly for the upper mass bin, these results are in excellent agreement with 
previous works. JOHN10, based on a sample of 1266 stars with \mstar $\sim$ 0.5\,-2.0 \msun, showed that the occurrence rate of planets 
increases linearly with the mass of the host star, reaching a fraction of $\sim$ 14 \% at \mstar $\sim$ 2.0 \msun. 
Similarly, based on a sample of 373 giant stars with \mstar $\sim$ 1.0\,-\,5.0 \msun, REF15 showed that the 
detection fraction of giant planets present a Gaussian distribution, with a peak in the detection fraction of $\sim$ 8\,\% at $\sim$ 1.9 \msun. 
Additionally, they showed that the occurrence rate around stars more massive than $\sim$ 2.7 \msun is consistent with zero, in good agreement with our 
findings, and also with theoretical predictions. For instance, based on a semi-analytic calculation of an evolving snow-line,
Kennedy \& Kenyon (\cite{KEN08}) showed that the formation efficiency increases linearly from 0.4 to 3.0 \msun. For stars more massive than 3.0 \msun, 
the formation of gas giant planets in the inner region of the protoplanetary disk is strongly reduced. Because of the fast stellar evolution timescale for
those massive stars, the snow line moves rapidly to 10-15 AU, preventing the formation of the giant planets in this region. \newline \indent
Figure \ref{occ_rate_metallicity} shows the planet occurrence rate as a function of the stellar metallicity. The symbols and lines are the 
same as in Figure \ref{occ_rate}. The width of the bins is 0.15 dex. 
It can be seen, that the occurrence rate increases with the stellar metallicity, with a peak of $f$ = 16.7$^{+15.5}_{-5.9}$\,\%
around stars with [Fe/H] = 0.35 dex. This trend seems to be real, despite a relatively high fraction observed in the bin centered at -0.25 dex, which might be 
explained by the low number statistics for that specific bin. Following the prescription of FIS05, we fitted the metallicity 
dependence of the occurrence fraction, with a function of the form:
\begin{equation}
     f([{\rm Fe/H}]) = \alpha\,10^{\beta\,[{\rm Fe/H}]}\,.
     \label{eq2}
\end{equation}
Using a similar approach for fitting equation \ref{eq1}, we obtained the following values: 
$\alpha$ = 0.061$^{+0.028}_{-0.003}$, and $\beta$ = 1.27$^{+0.83}_{-0.42}$ dex$^{-1}$. 
Figure \ref{alpha_beta} shows the probability density distribution of $\alpha$ and $\beta$ obtained after fitting the synthetic datasets.
The functional dependence of the occurrence rate with [Fe/H] (equation \ref{eq2}) is overplotted (solid curve). \newline \indent 
This relationship between the occurrence rate and the stellar metallicity is also observed in
solar-type stars (Gonzalez \cite{GON97}; Santos et al. \cite{SAN01}). Moreover, according to REF15, this trend is also present in giant stars. 
Interestingly, they also showed that there is an overabundance of planets around giant stars with [Fe/H] $\sim$ -0.3, similarly
to what we found in our sample. \newline  \indent
To investigate whether one of the two correlations presented above are spurious, we investigated the level of correlation between the stellar
mass and metallicity in our sample. Figure \ref{mass_fe_H} shows the mass of the star as a function of the metallicity for all of our targets (filled dots).
The open circles are the planet-hosting stars. In the top left corner is shown the mean uncertainty in [Fe/H] and \mstar.
From Figure \ref{mass_fe_H}, it is clear that there is some dependence between these two quantities. The Pearson linear coefficient is r\,=\,0.27, 
which means that there is an insignificant level of correlation. Moreover, if we restrict our analysis to stars with \mstar $<$ 2.5 \msun, the r-value drops
to 0.22, and we obtain a steeper rise of the occurrence rate with the stellar metallicity. Thus, we conclude that the two correlations 
presented in Figures \ref{occ_rate} and \ref{occ_rate_metallicity} are valid.  \newline \indent
In addition, we computed the fraction of our stars hosting planets in the stellar mass-metallicity space, using the same bins size presented in
REF15. These results are listed in Table \ref{det_frac}. Columns 1 and 2 correspond to the stellar metallicity and mass bins, each of 0.16 dex and 0.8 
\msun, respectively. The number of stars with detected planets (n$_p$), number of stars in the bin (n$_s$) and the fraction of stars with planets in 
each bin ($f$), are listed in columns 3-5. 
It can be seen that the highest fraction is obtained in the bin centered at 1.4 \msun and 0.12 dex ($f$\,=\,9.4$^{+ 8.3}_{- 5.1}$), which is 
slightly higher than the value of the bin with the same metallicity, but centered at 2.2 \msun ($f$\,=\,9.1$^{+10.8}_{- 5.8}$). Interestingly,
REF15 found a similar trend, i.e., they also obtained the highest fraction in these two mass-metallicity bins, although they claim higher values for $f$.
Also, we analyzed the combined results of the two surveys. These are listed in columns 6-8\, in Table \ref{det_frac}. It can be seen that 
the overall trend is unaffected, but the uncertainties in the planet fraction are smaller. \newline \indent
Finally, to understand whether the combined results are affected by systematic differences in the stellar parameters derived independently by the two surveys, 
we compared the resulting metallicities of the Lick Survey (listed in REF15) with those derived using our method. We used a total of 16 stars, from which 12 of them are 
common targets. We measured a difference of $\Delta$([Fe/H]) = 0.03 dex $\pm$ 0.11 dex, showing the good agreement between 
the two methods\footnote{Our typical dispersion (RMS) in the metallicity derived by each of the $\sim$ 150 Fe\,{\sc i} individual lines is $\sim$ 0.1 dex, corresponding
to an internal error $\lesssim$ 0.01 dex.}. Similarly, we compared the masses of the stars derived by the two surveys. We found that our stellar masses are 
on average larger by $\Delta$(\mstar) = 0.15 $\pm$ 0.37 \msun, which corresponds to a ratio of 1.07 $\pm$ 0.18.  
For comparison, Niedzielski et al. (\cite{NIE16}) also found that our stellar masses are overestimated with respect to their values by a factor 1.15 $\pm$ 0.10.
We also note that the planet-hosting star HIP\,5364 is a common target of the two surveys. They obtained \teff = 4528 $\pm$ 19 K and [Fe/H] = 0.07 $\pm$ 0.1 dex.
Using these values they derived a mass of 1.7 
$\pm$ 0.1 \msun for this star (Trifonov et al. \cite{TRI14}), significantly lower than our value of 2.4 $\pm$ 0.3 \msun. This shows that the combined results
of the Lick and EXPRESS surveys should be taken with caution. 
Certainly, a more detailed comparison between the stellar parameters derived independently by the two surveys, as well as their completeness, will allow us to 
check the validity of these combined results.
%
%

\begin{table*}
\centering
\caption{Detection fraction in different stellar mass-metallicity bins. \label{det_frac}}
\begin{tabular}{cccccccc}
\hline\hline
\vspace{-0.3cm} \\
                &           &           &   EXPRESS  &                       &        & EXPRESS + LICK &        \\
\cmidrule(lr){3-5} \cmidrule(lr){6-8}
${\rm [Fe/H]}$  & M$_\star$ &  n$_p$    &    n$_s$   &  $f$                  &  n$_p$ &   n$_s$        &  $f$   \\
      (dex)     &  (\msun)  &           &            &  (\%)                 &        &                &  (\%)  \\
\hline
\vspace{-0.3cm} \\
  -0.20         &   1.4     &     1     &    17      & 5.9$^{+11.3}_{-1.9}$  &  1     &     58     & 1.7$^{+3.8}_{-0.5}$ \\\vspace{-0.2cm} \\
  -0.20         &   2.2     &     0     &     5      & 0.0$^{+26.4}_{-0.0}$  &  2     &     34     & 5.9$^{+6.9}_{-1.9}$ \\\vspace{-0.2cm} \\
  -0.20         &   3.0     &     0     &     0      & -\,-\,-               &  0     &     21     & 0.0$^{+8.0}_{-0.0}$ \\\vspace{-0.2cm} \\
  -0.04         &   1.4     &     1     &    25      & 4.0$^{+ 8.1}_{- 1.3}$ &  3     &     54     & 5.6$^{+4.9}_{-1.7}$ \\\vspace{-0.2cm} \\
  -0.04         &   2.2     &     1     &    19      & 5.3$^{+10.3}_{- 1.7}$ &  2     &     70     & 2.9$^{+3.6}_{-0.9}$ \\\vspace{-0.2cm} \\
  -0.04         &   3.0     &     0     &     1      & 0.0$^{+60.2}_{- 0.0}$ &  0     &     30     & 0.0$^{+5.8}_{-0.0}$ \\\vspace{-0.2cm} \\
  +0.12         &   1.4     &     3     &    32      & 9.4$^{+ 7.8}_{- 3.0}$ &  7     &     48     & 14.6$^{+6.5}_{-3.7}$ \\\vspace{-0.2cm} \\
  +0.12         &   2.2     &     2     &    22      & 9.1$^{+ 9.9}_{- 3.1}$ &  6     &     46     & 13.0$^{+6.6}_{-3.5}$ \\\vspace{-0.2cm} \\
  +0.12         &   3.0     &     0     &    14      & 0.0$^{+11.5}_{- 0.0}$ &  2     &     36     & 5.6$^{+6.5}_{-1.8}$\\\vspace{-0.2cm} \\
\vspace{-0.3cm} \\\hline\hline
\end{tabular}
\end{table*}

%
%
%

\subsection{Multiple-planet systems.}

Out of the 12 planet-hosting stars in our sample, HIP\,5364, HIP\,24275 and HIP\,67851 host planetary systems with at least two 
giant planets. This means that 25\% of the parent stars, host a multiple system. Considering the full sample, it yields a 
$\sim$\,2\,\% fraction of multiple systems, comprised by two or more giant planets (M$_{\rm p}>$ 1.0\,\mjup). This number is a lower limit, since 
there are several other systems in our sample whose velocities are compatible with the presence of a distant giant planet, but still need 
confirmation (e.g. HIP\,74890, presented in section \ref{sec_HIP74890}).  \newline \indent
If we consider all of the known planet-hosting giant stars (log$g$ $\lesssim$ 3.6), around 10\% 
of them host a planetary system comprised by at least two giant planets. This fraction is significantly higher compared to solar-type stars. 
In fact, there are only 21 such systems among dwarf stars\footnote{source: http://exoplanets.org}, 
despite the fact that most of the RV surveys have targeted those type 
of stars. Moreover, planets are easier to be detected via precision RVs around solar-type stars, because they are on average less massive and 
have p-modes oscillations much weaker than giant stars (Kjeldsen \& Bedding \cite{KJE95}), 
which translates into larger amplitudes with a lower level of RV noise.
This observational result is a natural extension of the known mass distribution of single-planet systems orbiting evolved stars, which is 
characterized by an overabundance of super-Jupiter-like planets (e.g. Lovis \& Mayor \cite{LOV07}; D\"ollinger et al. \cite{DOL09}; 
Jones et al. \cite{JON14}). This result also reinforces the observed positive correlation between the stellar and planetary mass, in the sense that
more massive stars not only tend to form more massive single planets, but also more massive multi-planet systems. 


\section{Summary and discussion}

In this work we present precision radial velocities of four giant stars that have been targeted by the EXPRESS project, during the past
six years. These velocities show periodic signals, with semi-amplitudes between $\sim$ 50\,-\,100 \ms, which are likely caused by the doppler 
shift induced by orbiting companions. We performed standard tests (chromospheric emission, line bisector analysis and photometric
variability) aimed at studying whether these RV signals have an intrinsic stellar origin. We found no correlation between the stellar intrinsic
indicator with the observed velocities. Therefore, we conclude that the most probable explanation of the periodic RV signals observed in these stars is the 
presence of substellar companions. 
The best Keplerian fit to the RV data of the four stars leads to minimum masses between \mplan = 2.4\,-\,5.5 \mjup and 
orbital periods $P$ = 562\,-\,1560 days. Interestingly, all of them have low eccentricities ($e$ $\le$ 0.16), confirming that
most of the giant planets orbiting evolved stars present orbital eccentricities $\lesssim$ 0.2 (Schlaufman \& Winn \cite{SCH13}; Jones et al. \cite{JON14}).
The RVs of HIP\,74890 also reveal the presence of a third object at large orbital separation ($a >$ 6.5 AU). The RV trend induced by this object is
most likely explained by a brown dwarf or a stellar companion. \newline \indent
We also present a statistical analysis of the mass-metallicity correlations of the planet-hosting stars in our sample. This sub-sample is comprised 
of 12 stars, drawn from a parent sample of 166 stars, which host a total of 15 giant planets.
We show that the fraction of giant planets $f$, increases with the stellar mass in the range between $\sim$ 1.0\,-\,2.1 \msun, despite the fact that planets are 
more easily detected around less massive stars. 
For comparison, we obtained $f$\,=\,2.6$^{+5.4}_{-0.8}$\,\% for \mstar $\sim$ 1.3\,\msun, and a peak of $f$\,=\,13.0$^{+10.1}_{-4.2}$\,\% for stars with \mstar 
$\sim$ 2.1 \msun. These results are in good agreement with previous works showing that the occurrence rate of giant planets exhibit a positive correlation with the
stellar mass, up to \mstar $\sim$ 2.0 \msun (e.g. JOHN10; REF15). For stars more massive than $\sim$ 2.5 \msun, 
the fraction of planets is consistent with zero. We fitted the overall occurrence distribution with a Gaussian function (see Eq. \ref{eq1}), obtaining
the following parameters: $C$ = 0.14 $^{+0.08}_{-0.01}$, $\mu$ = 2.29 $^{+0.44}_{-0.06}$ \msun, and $\sigma$ = 0.64 $^{+0.44}_{-0.03}$ \msun. \newline \indent
Similarly, we studied the occurrence rate of giant planets as a function of the stellar metallicity. 
We found an overabundance of planets around metal-rich stars, with a peak of $f$ = 16.7$^{+15.5}_{-5.9}$\,\% for stars with [Fe/H]\,$\sim$\,0.35 dex. 
We fitted the metallicity dependence of the occurrence rate with a function of the form $f = \alpha10^{\beta\,[{\rm Fe/H}]}$, 
obtaining the following parameter values: $\alpha$ = 0.061$^{+0.028}_{-0.003}$, and $\beta$ = 1.27$^{+0.83}_{-0.42}$ dex$^{-1}$.
Our power-law index $\beta$ lies in between the values measured by JOHN10 ($\beta$ = 1.2 $\pm$ 0.2) and FIS05 ($\beta$ = 2.0).
Thus, our results suggest that the planet-metallicity correlation observed in solar-type stars is also present in 
intermediate-mass (\mstar $\gtrsim$ 1.5 \msun) evolved stars, in agreement with REF15 results. \newline \indent
Finally, we investigated the fraction of multiple planetary systems comprised by two or more giant planets. Out of the 12 systems presented above, three of them
contain two giant planets, which is a significant fraction of the total number of these planetary systems. If we also consider multi-planet systems published by other 
RV surveys, we found that there is a significantly higher fraction of them around intermediate-mass evolved stars in comparison to solar-type stars. This result is 
not surprising, since different works have shown that giant planets are more frequent around intermediate-mass stars (D\"ollinger et al. \cite{DOL09}; 
Bowler et al. \cite{BOW10}), which is also supported by theoretical predictions (Kennedy \& Kenyon \cite{KEN08}). Also, 
planets tend to be more massive around intermediate-mass stars compared to those around solar-type stars (e.g. Lovis \& Mayor \cite{LOV07}; D\"ollinger et al. \cite{DOL09}; 
Jones et al. \cite{JON14}). Thus, we conclude that the high fraction of multiple systems observed in giant stars is a natural consequence of the planet 
formation mechanism around intermediate-mass stars.


\begin{acknowledgements}
M.J. acknowledges financial support from Fondecyt project \#3140607 and FONDEF project CA13I10203.
A.J. acknowledges support from Fondecyt project \#1130857, the Ministry
of Economy, Development, and Tourism's Millennium Science Initiative
through grant IC120009, awarded to The Millennium Institute of
Astrophysics (MAS). P.R acknowledges funding by the Fondecyt project \#1120299.
F.O acknowledges financial support from Fondecyt project \#3140326 and from the MAS.
J.J., P.R and A.J acknowledge support from CATA-Basal PFB-06. 
HD acknowledges financial support from FONDECYT project \#3150314.
This research has made use of the SIMBAD database and the VizieR catalogue access tool, operated at CDS, Strasbourg, France. 
\end{acknowledgements}

\Online

\begin{appendix} 

\section{Radial velocity tables.}


%
\begin{table}
\centering
\caption{Radial velocity variations of HIP\,8541\label{HIP8541_vels}}
\begin{tabular}{lccc}
\hline\hline
JD\,-\,2450000        & RV & error  & Instrument\\
          &  (\ms)   &   (\ms) \\
\hline \vspace{-0.3cm} \\
  5457.7875  & -136.5 & 5.1 & FEROS \\
  6099.9286  &   59.6 & 4.9 & FEROS \\
  6110.8570  &   41.3 & 3.8 & FEROS \\
  6160.8413  &   25.2 & 3.5 & FEROS \\
  6230.6666  &   23.2 & 4.1 & FEROS \\
  6241.7042  &   23.2 & 3.8 & FEROS \\
  6251.7343  &   24.4 & 2.9 & FEROS \\
  6321.5772  &    1.4 & 4.2 & FEROS \\
  6565.7347  &  -61.8 & 4.7 & FEROS \\
  6533.8456  &  78.4  & 5.1& CHIRON \\
  6823.9189  &   2.9  & 5.1& CHIRON \\
  6836.8384  &  17.3  & 5.1& CHIRON \\
  6850.7623  &   0.6  & 4.8& CHIRON \\
  6882.7614  &  -0.6  & 4.0& CHIRON \\
  6909.8364  & -15.4  & 4.0& CHIRON \\
  6939.5587  & -26.2  & 4.1& CHIRON \\
  6958.5726  & -30.7  & 4.3& CHIRON \\
  6972.6441  & -25.9  & 4.4& CHIRON \\
  6993.5329  & -37.6  & 3.6& CHIRON \\
  7017.6540  & -11.2  & 4.0& CHIRON \\
  7060.5258  & -39.9  & 4.3& CHIRON \\
  7070.5331  &  -9.6  & 4.7& CHIRON \\
  7080.5042  & -25.0  & 6.6& CHIRON \\
  7163.9265  & -17.9  & 6.0& CHIRON \\
  7192.8795  & -14.2  & 4.3& CHIRON \\
  7253.7358  &  -0.3  & 4.9& CHIRON \\
  7273.8077  &  10.0  & 3.9& CHIRON \\
  7284.8938  &  20.1  & 4.9& CHIRON \\
  7286.8216  &  32.6  & 5.2& CHIRON \\
  7293.7664  &  27.6  & 4.6& CHIRON \\
  7311.7297  &  25.8  & 3.9& CHIRON \\
  7332.6564  &  39.3  & 3.8& CHIRON \\
  5138.1360  & -53.1  & 1.7 & UCLES  \\
  5496.1501  & -94.2  & 1.8 & UCLES  \\
  5525.9950  & -83.6  & 1.8 & UCLES  \\
  5880.0887  &  11.4  & 2.3 & UCLES  \\
  5969.9464  &  44.3  & 2.3 & UCLES  \\
  6527.2141  &  0.00  & 2.2 & UCLES  \\
\hline \vspace{-0.3cm} \\
\hline\hline
\end{tabular}
\end{table}

\begin{table}
\centering
\caption{Radial velocity variations of HIP\,74890\label{HIP74890_vels}}
\begin{tabular}{lccc}
\hline\hline
JD\,-\,2450000        & RV & error  & Instrument\\
          &  (\ms)   &   (\ms) \\
\hline \vspace{-0.3cm} \\
   5317.7161  &   43.8 & 5.3 & FEROS \\
   5336.8361  &   39.7 & 4.4 & FEROS \\
   5379.7325  &   41.7 & 3.9 & FEROS \\
   5428.5941  &   41.8 & 4.9 & FEROS \\
   5729.7413  &   68.9 & 4.8 & FEROS \\
   5744.6796  &   82.7 & 5.1 & FEROS \\
   5793.6145  &   59.0 & 6.2 & FEROS \\
   6047.6958  &  -12.2 & 2.7 & FEROS \\
   6056.6979  &  -31.1 & 2.9 & FEROS \\
   6066.6960  &  -24.8 & 3.8 & FEROS \\
   6099.6861  &  -21.7 & 3.5 & FEROS \\
   6110.6576  &  -17.6 & 4.2 & FEROS \\
   6140.6536  &  -28.7 & 4.9 & FEROS \\
   6321.8844  &  -28.2 & 4.0 & FEROS \\
   6342.9030  &  -16.0 & 3.8 & FEROS \\
   6412.6527  &  -10.4 & 3.0 & FEROS \\
   7114.8393  & -102.5 & 4.2 & FEROS \\
   7174.5725  &  -84.3 & 5.2 & FEROS \\
   4869.2727  &  158.3 & 2.4 & UCLES \\
   5381.0725  &   27.9 & 1.9 & UCLES \\
   5707.0809  &   77.6 & 4.6 & UCLES \\
   5969.2830  &   22.7 & 1.5 & UCLES \\
   5994.1579  &    5.6 & 2.0 & UCLES \\
   6052.0874  &  -24.0 & 3.6 & UCLES \\
   6088.9942  &  -22.0 & 1.7 & UCLES \\
   6344.2382  &  -13.0 & 1.9 & UCLES \\
   6375.2713  &   -0.0 & 1.9 & UCLES \\
   6400.1449  &    0.0 & 1.7 & UCLES \\
   6494.9152  &    1.3 & 2.0 & UCLES \\
   6529.9054  &   -2.3 & 2.4 & UCLES \\
   6747.1435  &  -37.8 & 1.6 & UCLES \\
\hline \vspace{-0.3cm} \\
\hline\hline
\end{tabular}
\end{table}

\begin{table}
\centering
\caption{Radial velocity variations of HIP\,84056\label{HIP84056_vels}}
\begin{tabular}{lccc}
\hline\hline
JD\,-\,2450000        & RV & error  & Instrument\\
          &  (\ms)   &   (\ms) \\
\hline \vspace{-0.3cm} \\
5379.7813  & -59.2 &  3.4 & FEROS \\
5428.6445  & -44.2 &  4.0 & FEROS \\
5457.5635  & -49.2 &  2.6 & FEROS \\
5470.5498  & -42.7 & 11.2 & FEROS \\
5744.7607  &  38.5 &  5.6 & FEROS \\
5786.6509  &  33.6 &  3.9 & FEROS \\
6047.7637  &  -2.7 &  7.1 & FEROS \\
6056.7446  &  -9.7 &  4.4 & FEROS \\
6066.7539  & -16.1 &  5.3 & FEROS \\
6099.7607  & -20.4 &  5.0 & FEROS \\
6160.6299  & -20.0 &  3.4 & FEROS \\
6412.8618  &  17.7 &  3.7 & FEROS \\
6431.7646  &  27.8 &  5.9 & FEROS \\
6472.6177  &  31.2 &  3.5 & FEROS \\
6472.6514  &   9.1 &  4.0 & FEROS \\
6472.6646  &  10.8 &  3.5 & FEROS \\
6472.7095  &  16.5 &  4.3 & FEROS \\
6472.7666  &  24.6 &  3.4 & FEROS \\
6565.5210  &  54.4 &  4.7 & FEROS \\
6722.8143  &  34.2 &  3.5 & CHIRON \\
6742.7472  &  43.1 &  3.5 & CHIRON \\
6743.7859  &  36.2 &  4.3 & CHIRON \\
6769.7072  &  37.5 &  3.5 & CHIRON \\
6785.7443  &  46.1 &  3.7 & CHIRON \\
6804.6654  &  20.6 &  4.3 & CHIRON \\
6822.6781  &   9.5 &  5.5 & CHIRON \\
6839.5918  &  14.2 &  3.7 & CHIRON \\
6885.5857  & -14.5 &  4.7 & CHIRON \\
6904.5370  & -16.0 &  5.3 & CHIRON \\
6919.5493  & -22.6 &  3.8 & CHIRON \\
6928.5084  & -18.4 &  4.0 & CHIRON \\
6929.5103  & -17.3 &  3.8 & CHIRON \\
6937.5205  & -28.3 &  4.1 & CHIRON \\
6944.4897  & -14.3 &  4.2 & CHIRON \\
6954.4887  & -28.4 &  4.1 & CHIRON \\
7088.8135  & -18.5 &  3.8 & CHIRON \\
7124.7610  & -29.7 &  4.0 & CHIRON \\
7145.9299  & -26.9 &  4.5 & CHIRON \\
7166.6458  & -15.0 &  3.9 & CHIRON \\
7177.7871  & -19.7 &  5.1 & CHIRON \\
7207.6342  & -13.4 &  4.6 & CHIRON \\
\hline \vspace{-0.3cm} \\
\hline\hline
\end{tabular}
\end{table}

\begin{table}
\centering
\caption{Radial velocity variations of HIP\,95124\label{HIP95124_vels}}
\begin{tabular}{lccc}
\hline\hline
JD\,-\,2450000        & RV & error  & Instrument\\
          &  (\ms)   &   (\ms) \\
\hline \vspace{-0.3cm} \\
   5379.8357  &   -1.0 & 5.1 & FEROS \\
   5457.6545  &   16.4 & 4.5 & FEROS \\
   5744.8034  &  -19.8 & 2.7 & FEROS \\
   5786.7273  &  -43.6 & 3.5 & FEROS \\
   5793.7489  &  -39.7 & 4.6 & FEROS \\
   6047.8077  &   50.0 & 3.6 & FEROS \\
   6056.7627  &   40.6 & 3.9 & FEROS \\
   6066.7758  &   44.6 & 3.9 & FEROS \\
   6110.7261  &   47.6 & 2.3 & FEROS \\
   6110.7330  &   48.0 & 1.9 & FEROS \\
   6110.7445  &   49.5 & 2.3 & FEROS \\
   6160.6791  &   35.5 & 3.9 & FEROS \\
   6241.5101  &    6.3 & 4.5 & FEROS \\
   6431.8159  &  -45.1 & 4.1 & FEROS \\
   6472.7252  &  -31.3 & 3.6 & FEROS \\
   6472.7816  &  -31.2 & 2.8 & FEROS \\
   6472.7947  &  -32.6 & 2.6 & FEROS \\
   6472.8171  &  -45.8 & 2.7 & FEROS \\
   6472.8474  &  -39.6 & 2.4 & FEROS \\
   6565.5720  &   -8.6 & 3.7 & FEROS \\
   6895.5348  &  -4.7  & 4.1 & CHIRON \\
   6909.6056  &  -6.3  & 3.9 & CHIRON \\
   6911.5434  & -15.1  & 5.2 & CHIRON \\
   6920.5681  & -11.2  & 3.9 & CHIRON \\
   6921.5119  & -19.9  & 4.1 & CHIRON \\
   6926.5602  & -15.8  & 3.8 & CHIRON \\
   6936.5852  & -16.0  & 4.6 & CHIRON \\
   6945.4849  & -12.0  & 6.9 & CHIRON \\
   6976.5018  & -26.2  & 4.2 & CHIRON \\
   7088.8888  &  27.0  & 4.1 & CHIRON \\
   7099.8966  &  18.1  & 5.2 & CHIRON \\
   5139.8852  &  4.7   & 4.9 & UCLES \\
   5396.9380  & -15.2  & 4.2 & UCLES \\ 
   6494.1188  & -14.2  & 2.3 & UCLES \\ 
   6528.9451  &  0.0   & 2.5 & UCLES \\ 
   6747.2858  &  26.5  & 2.2 & UCLES \\ 
\hline \vspace{-0.3cm} \\
\hline\hline
\end{tabular}
\end{table}


\end{appendix}

\end{document}